\pdfoutput=1

\documentclass[11pt]{article}

\usepackage{acl}

\usepackage{times}
\usepackage{latexsym}

\usepackage[T1]{fontenc}

\usepackage[utf8]{inputenc}

\usepackage{microtype}

\usepackage{inconsolata}

\usepackage{graphicx}

\usepackage{geometry}
\usepackage{array}
\usepackage{booktabs}
\usepackage{pdflscape}
\newcommand{\xhdr}[1]{\vspace{1mm}\noindent{{\bf #1.}}}
\usepackage[utf8]{inputenc}

\newcommand{\remove}[1]{}

\title{Measuring Mental Health Variables in Computational Research:\\Toward Validated, Dimensional, and Transdiagnostic Approaches}

\author{Chen Shani\\
  Computer Science\\
  Stanford University\\
  \href{mailto:cshani@stanford.edu}{cshani@stanford.edu} 
  \And
  Elizabeth C. Stade \\
  Human-Centered Artificial Intelligence\\
  Stanford University\\
  \href{mailto:ecs@stanford.edu}{ecs@stanford.edu}
  }
\begin{document}
\maketitle
\begin{abstract}

Computational mental health research develops models to predict and understand psychological phenomena, but often relies on inappropriate measures of psychopathology constructs, undermining validity. We identify three key issues: (1) reliance on unvalidated measures (e.g., self-declared diagnosis) over validated ones (e.g., diagnosis by clinician); (2) treating mental health constructs as categorical rather than dimensional; and (3) focusing on disorder-specific constructs instead of transdiagnostic ones. We outline the benefits of using validated, dimensional, and transdiagnostic measures and offer practical recommendations for practitioners. Using valid measures that reflect the nature and structure of psychopathology is essential for computational mental health research.
\end{abstract}

\section{Introduction}
In computational mental health research, significant effort is invested in designing models to predict and understand psychological phenomena. Yet the validity and utility of these models can be undermined when they rely on flawed or inappropriate representations of psychopathology (i.e., mental disorder). For example, a classifier cannot validly predict depression if its training data is based on an invalid measure of depression.

Measuring psychopathology \textit{constructs} -- representations of psychological states or processes like ``depression'' or ``neuroticism'' \cite{fried_what_2017} -- is challenging. Constructs are abstract and cannot be directly observed (e.g., there is no single biological indicator for depression) and diagnostic systems like \textit{DSM-5} \cite{american2013diagnostic} and \textit{ICD-10} \cite{whoicd10} have slight variations in how they define syndromes. 

Yet clinical psychology has for decades invested in validating  measures of psychopathology \cite{cronbach_construct_1955} and advancing measurement techniques that reflect developments in psychological science \cite{stanton2020transdiagnostic}. Recommended measurement techniques include clinician-administered interviews, self-report questionnaires, and informant reports \cite{stanton2020transdiagnostic}. While no measurement scheme is without error, techniques exist to help ensure that a measure actually taps the construct it purports to (a process known as \textit{construct validity} (\citet{campbell_convergent_1959}).


However, computational psychopathology research has slow to adopt measurement techniques from clinical psychological science, hindering progress. Here, we highlight three key concepts from clinical science—measurement validity, dimensional measurement, and transdiagnostic measurement. Acknowledging that
computational research often turns to inappropriate measures of psychopathology constructs due to the constraints of computational research -- collecting large-scale clinician-assessed data is expensive and time-consuming, and existing or archival data can be difficult or impossible to access due to privacy concerns -- we offer practical recommendations for improved assessment of psychological phenomena in computational research.

\section{Measurement Validity}
Computational research often infers mental health conditions using methods with poor or unknown validity. For example, some studies assume a diagnosis based on forum membership (e.g., \texttt{r/Depression}) or self-declarations (e.g., ``I have depression'') on social media \cite{guntuku_detecting_2017}, ``proxy diagnostic signals'' which have been shown to have poor external validity \cite{ernala2019methodological}. In other cases, computational researchers write their own single-item measure of a psychological construct, rather than selecting an existing measure with good validity \cite{allen2022single}. The clinical science literature provides justification that all measurement methods are not equal vis-a-vis validity: For instance, self-reported history of depression diagnosis only modestly agrees with semi-structured diagnostic interview findings \cite{stuart2014comparison, sanchez2008validity}. 


To mitigate this, computational researchers should use measures of psychological constructs with good validity. Gold standard psychopathology assessment typically involves a clinician-administered structured or semi-structured interview (e.g., Structured Clinical Interview for \textit{DSM-5}; \citet{first_scid_2016} based on an established psychopathology classification system (e.g., \textit{DSM-5}. As an alternative, self- or informant-report measures that have undergone rigorous psychometric evaluation, such as the PHQ-9 \cite{kroenke1999phq9}, can be used \cite{stanton2020transdiagnostic}.

\subsection{Limitations of Self-Report Questionnaires}

Self-report methodology, while offering high levels of convenience,  has meaningful limitations of which computational researchers should be aware:

\xhdr{Bias} Self-report measures are vulnerable to biased patterns of responding such as participant lack of insight or yay- or nay-saying biases \cite{hunt2003self}, which can introduce systematic errors into computational models.

\xhdr{Specificity of Constructs} Self-report tools may be imprecise measures of psychological constructs. Evidence exists that putative self-report measures of depression may in fact capture general distress \cite{coyne1994self, kendall1987issues} or anxiety \cite{breslau1985depressive} rather than just depression. This lack of precision can weaken model predictions and blur construct boundaries.

\xhdr{Using Tools Outside Their Intended Setting} Self-report measures may be less effective when used outside their original context. For example, the PHQ-9, designed for primary care, has low specificity and PPV in specialty mental health settings \cite{inoue2012utility}, limiting its validity in clinical samples. Developed as a screening tool, it maps perfectly onto \textit{DSM-5} criteria, but it does not assess other symptoms associated with depression, like self-dislike or low libido, meaning it has a narrower possible range than measures that capture a wide range of symptoms. Using tools beyond their intended purpose may reduce model accuracy, and range restriction can attenuate effects (see below).


\section{Dimensional Measurement}

Mental health constructs can be assessed dimensionally (e.g., on a scale from 0 [\textit{no depression}] to 10 [\textit{extreme depression}]) or using categorical labels (\textit{not depressed} vs. \textit{depressed}). Despite computational research tending to employ categorical measurement schemes, most psychopathology constructs are inherently dimensional, as evidenced by the following:

At the \textbf{manifest (observable) level}, symptoms show a monotonic relationship with functional outcomes (\citet{kessler_epidemiology_2006, ruscio_broadening_2007, ruscio_social_2008, cuijpers2004minor, judd1997role}). Even mild or infrequent depression symptoms that fall below the \textit{DSM-5} criteria for major depressive disorder are associated with impairment. However, categorical representations of psychopathology group all subthreshold symptom presentations together, obscuring mild yet clinically meaningful dysfunction \cite{ruscio2019normal}.

Furthermore, longitudinal research reveals that individuals frequently fluctuate between levels of severity of symptoms over time, including crossing thresholds above and below thresholds week-by-week \cite{chen2000empirical, judd1997role}, making diagnoses somewhat arbitrary depending on the time of evaluation.

At the \textbf{latent level}, taxometric analysis of psychometric variables, which compares categorical and dimensional models \cite{ruscio2000informing}, typically yields dimensional solutions. This indicates that constructs like depression and anxiety have a natural, underlying structure that is dimensional, not categorical \cite{haslam_categories_2012}.

To reflect the true nature and structure of psychopathology, computational researchers should treat most mental health variables as continuous. Using dimensional measures of psychopathology can improve granularity of computational models (e.g., allowing models to differentiate between moderate and severe symptom presentations), improving accuracy and clinical utility.

\section{Transdiagnostic Measurement}

While diagnostic systems like the \textit{DSM-5} dominate clinical practice and research, there is growing recognition of \textit{transdiagnostic processes} that cut across the mental disorders \citep{harvey2004cognitive}. For example, \textit{avoidance} is shared by many different types of anxiety disorders, including panic disorder, specific phobia, and social anxiety disorder. Growing evidence supports transdiagnostic conceptualizations of psychopathology, including:

\xhdr{High Comorbidity} Mental disorders frequently co-occur at rates far exceeding chance. For example, people diagnosed with one anxiety disorder are six times more likely to have another \cite{kessler-1997-comorbidity}, suggesting shared underlying mechanisms.

\xhdr{High Diagnostic Crossover} Many patients transition between diagnoses over time. For example, 20-50\% of individuals diagnosed with anorexia nervosa later develop bulimia \cite{eddy2008diagnostic}, suggesting that existing diagnostic categories may sub-optimally represent psychopathology.

\xhdr{Non-Specific Treatment Effects} Treatments targeting one disorder often alleviate symptoms of co-occurring conditions—for example, PTSD treatments frequently reduce depression symptoms \cite{barlow2014nature}. This suggests that interventions may be acting on transdiagnostic mechanisms rather than disorder-specific factors.

In response, the field has introduced transdiagnostic classification models, such as \textbf{NIMH's Research Domain Criteria (RDoC)} \cite{insel2010research}, which defines cross-cutting dimensions like ``reward responsiveness'' and "potential threat," and the \textbf{Hierarchical Taxonomy of Psychopathology (HiTOP)} \cite{kotov2017hierarchical}, which groups together frequently co-occurring symptoms. 



Transdiagnostic measures offer increased parsimony and can better reflect the psychopathology vis-\`a-vis syndrome-specific alternatives (e.g., \citet{stade2023transdiagnostic}; See \citet{stanton2020transdiagnostic} for guidance on selecting transdiagnostic measures).

\section{Impact of Poor Measurement}
Using measures with poor validity, or using categorical or syndrome-specific measures, can negatively impact computational research. Key implications include:

\xhdr{Reduced Resolution} Using categorical labels for psychopathology oversimplifies complex constructs, discarding valuable information about symptom severity, which is important for model accuracy and has clinical utility (e.g., the difference between moderate and severe depression is meaningful to clinicians).

\xhdr{Mis-Classifying Boundary Cases} Relatedly, categorical representation of psychopathology risks misclassifying individuals who fall close to the diagnostic boundary. 

\xhdr{Risk of Type II Errors} Dichotomizing variables that are continuous in nature sacrifices statistical power \cite{cohen1983cost} and reduces reliability \citep{markon2011reliability}, increasing the risk of Type II errors.

\xhdr{Overfitting and Poor Generalization} Noisy measurements cause models to learn spurious patterns, reducing reliability and real-world applicability.

\xhdr{Misleading Interpretations} Poor measurements can cause misleading conclusions about mental health constructs, such as conflating depression overlapping yet different constructs, like anxiety or negative emotionality.

\xhdr{Erosion of Clinical Utility} For computational models to have practical relevance in mental health care, they must provide insights or predictions that clinicians can act upon. Models based on bad measurements often lack this clinical utility. 

\xhdr{Bias Amplification and Inequities} Inaccurate measurement can amplify bias, reinforcing disparities and inequities in mental health care.

\xhdr{Missed Opportunities for Scientific Progress} Bad measurements limit scientific progress, preventing meaningful contributions and advancements in understanding mental health.

\section{The Path Forward}

By addressing the limitations of current measurement practices in computational mental health research, we hope to create more accurate, robust, and impactful models. To strengthen the scientific rigor and relevance of research, we offer the following recommendations:

\xhdr{Consult Experts} Computational researchers can refer to established guidelines and evidence-based recommendations for the assessment of specific constructs or disorders (e.g., \citet{klein2005toward, antony2005evidence, shear1994standardized}). Collaborating with clinical science colleagues across departments can help guide appropriate measure selection. For projects aiming at immediate clinical application, working closely with a clinician is essential. Clinicians can offer expertise on the tool's clinical utility, including its relevance to real-world practice, ease of use in clinical settings, and alignment with existing diagnostic and treatment workflows. They can also provide feedback on whether the tool offers actionable insights for patient care, supports case conceptualization and treatment planning, and meets the practical needs of diverse clinical populations.


\xhdr{Improve Methodological Rigor When Using Proxies} As previously described, proxy diagnostic signals, such as self-identified diagnoses on social media, have poor validity \cite{ernala2019methodological}. Suggestions for improving the methodological rigor of research using proxies include paring proxy diagnostic signals with offline clinical datasets \cite{inkster2016decade} -- a strong correlation between the proxy and an established clinical outcome, even in just a subsample of participants, could serve to demonstrate the validity of the proxy variable --  and combining multiple proxies to improve reliability \cite{ernala2019methodological}. At the very least, researchers using proxies should clearly state their limitations, e.g., a Twitter-based depression variable should be distinguished from a clinically validated diagnosis, with a note that future research using higher validity measures is needed. 

\xhdr{Adopt Dimensional Measurement} Avoid measuring mental health constructs into binary categories (e.g., ``depressed'' vs. ``not depressed''). Choose dimensional measures that capture severity gradients and avoid dichotomizing continuous variables to form diagnostic categories. Researchers interested in diagnostic status could test this variable in secondary analyses (e.g., \cite{stade2023depression}).

\xhdr{Critically Evaluate Disorder-Specific Measures} Before selecting a construct of interest and its corresponding measure, carefully evaluate whether a disorder-specific approach is necessary. For example, many researchers express interest in indexing anxiety, yet do so using the GAD-7 \cite{spitzer2006brief}, which measures the symptoms of generalized anxiety disorder, a disorder of frequent and uncontrollable worry. Yet uncontrollable worry only represents one form of anxiety pathology. Measures that are not disorder-specific (e.g., MASQ Anxious Symptoms subscale \cite{watson1995testing}) better capture features and processes that cut across the anxiety disorders.  Researchers should use syndrome-specific measures only when this truly aligns with their research goals.

\xhdr{Adopt a Process-Oriented Approach} Instead of focusing solely on specific disorders or syndromes, consider examining transdiagnostic processes. For example, studying constructs that encompass multiple diagnostic categories, such as ``internalizing psychopathology'' or ``fear," can offer more generalizable and integrative insights than research limited to a single diagnosis (e.g., major depression, specific phobia). 


\xhdr{Think Beyond Psychopathology} There are many non  \textit{DSM-5} constructs that are important for health and well-being, especially those that confer risk or protection for psychopathology, such as neuroticism, perfectionism, resilience, and disinhibition. The HiTOP ``components/traits'' level of analysis (e.g., \citet{deyoung2022distinction} offers a starting place for exploring non \textit{DSM-5} constructs.

\xhdr{Maximize Range on Variables of Interest} Since many psychopathology constructs are dimensional, researchers should recruit participants with varying levels of the construct. For example, when studying depression, aim to include the widest possible range of depression severity scores, including subthreshold presentations (e.g., individuals who have symptoms of depression that do not meet \textit{DSM-5} major depressive disorder criteria). Maximizing range on variables of interest should also yield greater effect sizes, since range restriction attenuates effects \cite{linn1968range}.

\xhdr{Attend to Reliability} Reliability sets the upper limit of validity and is crucial for research. Assess reliability using metrics like Cronbach's alpha for self-report \cite{cronbach1951coefficient}, the intraclass correlation coefficient, for dimensional observer ratings, or Cohen's Kappa, for categorical observer ratings \cite{hallgren2012computing}. Training raters thoroughly and enhancing rater competency can ensure good reliability \cite{reichelt2003impact, creed2016beyond}.

\xhdr{Consider Condition Heterogeneity} Mental health conditions are highly heterogeneous, with significant variability in symptom presentation and individual experiences. Two patients with a \textit{DSM-5} diagnosis of major depressive disorder may not share a single symptom \cite{fried2017dep:heterogeneity}. Study designs should account for this variability -- including by analyzing individual symptoms \cite{fried2015depression} -- to avoid oversimplification.

\xhdr{Address Comorbidity} Mental health conditions often co-occur and share overlapping symptoms, hinting that effects thought to be driven by one disorder could be driven by co-occurring conditions. Researchers can account for comorbidity using statistical controls (e.g., \cite{stade2023depression}) or measures that disentangle overlapping conditions (e.g., \cite{watson1995testing}), increasing confidence that effects are unique to a given condition. 

\xhdr{Adopt Longitudinal Measurement} Psychopathology dynamically evolves over time, both in terms of severity and diagnostic label. Longitudinal methods of data collection, and analyses using temporally-aware models or time-series analyses, could help address this reality. Although not historically accounted for in computational research, recent work has begun to examine the relationship between symptoms and language features over time (e.g., \citet{nook2022linguistic}).


We acknowledge that some of the proposed suggestions are challenging for NLP researchers to implement: Computational researcher may require sample sizes prohibitively large for conducting semi-structured diagnostic interviewing (prohibitive from a resource perspective, and even the interaction required to collect self-reported scores, as opposed to social-media based proxy measurement, which may require no interaction between researchers and participants, may be more resource-intensive or involved than is actually feasible. Especially if doing social-media based research requires no interaction with participants whatsoever. Therefore, while advocating for clinician-assessed, dimensional psychopathology measurement as the gold-standard, we suggest that researchers seeking to strengthen measurement approaches can adopt an ``n+1'' approach, where they seek to take one step towards improved measurement. For example, researchers planning to administer a single-item measure can weigh the pros and cons of this approach \cite{allen2022single} and select a measure with demonstrated validity in their population of interest (e.g., \citet{joiner2025reliability}) rather than writing their own item from scratch. Researchers can follow the guidelines for selecting measures in line with transdiagnostic frameworks (e.g. \citet{stanton2020transdiagnostic}) rather than using disorder-specific measures. To demonstrate what different measurement strategies can look like, we present in Table \ref{tab:anxiety-measures} a matrix demonstrating measures of social anxiety that systematically vary on the categories we have highlighted in this paper (validated vs. unvalidated, categorical vs. dimensional, and disorder-specific vs. transdiagnostic). Even incremental improvements can significantly improve validity and utility.

Beyond this, the field sorely needs large, publicly available datasets that include natural language from well-characterized clinical samples, perhaps created leveraging something like a practice-research network \cite{parry2010practice}. Given that many academic mental health clinics routinely administer the same semi-structured interviews, the aggregation of such recordings could be utilized. Diagnostic interviews can be a particularly efficient source of data, because they yield language as well as measurement of psychopathology constructs, and they are often audio recorded. It is possible to use to predict diagnostic severity scores obtained from a separate section of the interview (e.g., \citet{stade2023depression}. 

Accruing this type of large, shared dataset is not without challenges, one of which is the issue of confidentiality. It is difficult to acquire natural language data that are not identifiable or semi-identifiable in some way; and to conduct computational research, this dataset would need clinical variables; risking the disclosure of PHI. However, a potential workaround is not making the raw language public but instead extracting a range of linguistic features (including basic, dictionary-based features as well as more sophisticated, transformer/embedding based features) available.

\section{Discussion and Conclusions}

We highlight challenges in measuring mental health constructs in computational research and propose ways to improve validity. Key issues include overreliance on categorical frameworks, neglect of condition heterogeneity, and inadequate transdiagnostic measures.

Categorical frameworks like the \textit{DSM-5} oversimplify constructs, while dimensional approaches—capturing severity and shared symptoms—enhance model accuracy. Focusing on transdiagnostic constructs, like ``negative affect,'' provides a holistic understanding of mental health.

Condition heterogeneity complicates analysis, but transdiagnostic approaches can address comorbidity and overlapping symptoms. Poor measurement practices introduce errors and biases, so researchers should prioritize validated instruments and diverse datasets.


We argue that researchers should adopt dimensional measures, assess disorder-specific metrics critically, and ensure sample diversity. Implementing transdiagnostic approaches, rater calibration, and reliability checks will further enhance validity.

In conclusion, improving measurement practices is crucial for advancing computational models and mental health care, capturing the complexity of psychopathology, and driving progress in the field.

\newpage
\section{Limitations}


While this work highlights critical issues in the measurement of mental health constructs and provides practical recommendations, it is important to acknowledge its limitations. 

First, although we emphasize the importance of validated measures, we recognize that resource constraints and practical barriers may prevent many researchers from employing clinician-rated assessments or developing fully validated instruments. These barriers underscore the need for scalable, cost-effective alternatives that balance feasibility and validity.

Second, while we center our discussion on practices and pitfalls in the computational research community, we do not mean to imply that computational researchers are the only ones making these mistakes. These measurement issues we outline here are common in behavioral health research more broadly, including psychiatric and psychological research. Addressing these challenges comprehensively will require a broader interdisciplinary effort that includes collaboration across fields.

Third, while we focus on dimensional and transdiagnostic measurement approaches, we acknowledge that these may not be universally applicable. Certain clinical scenarios may necessitate categorical diagnoses for treatment decisions, and some researchers may have justifiable reasons for focusing on specific disorders. Future work should aim to provide clearer guidance on when categorical, dimensional, or transdiagnostic approaches are best.

Fourth, this work does not claim to exhaustively address all the challenges in the measurement of mental health constructs. Other significant issues, such as the influence of cultural biases, ethical considerations in mental health data collection, and the challenges of interpreting results from large-scale datasets, also warrant attention but fall outside the scope of this discussion. 

Finally, while we provide practical recommendations, the field still lacks consensus on "best practices" for measuring mental health constructs in computational research. More empirical studies are needed to evaluate the relative merits of different measurement approaches and their impacts on model performance and real-world applications.

Despite these limitations, we hope that this work stimulates critical reflection and contributes to advancing the validity and utility of mental health research in computational contexts.

\section{Acknowledgements}
Dan Jurafsky  provided critical feedback on an earlier version of this manuscript, and a conversation with Ayelet M. Ruscio shaped our thinking about the different measurement strategies appearing in Table \ref{tab:anxiety-measures}.
This work was supported in part by the Koret Foundation grant for Smart Cities and Digital Living.

\bibliography{custom}

\newpage

\section{Appendix}
\label{sec:appendix}

\begin{table*}[htbp]
\centering
\rotatebox{90}{
\begin{tabular}{|p{2.3cm}|p{4.1cm}|p{4.1cm}|p{4.1cm}|p{4.1cm}|}
\hline
& \multicolumn{2}{c|}{\textbf{Not dimensional}} & \multicolumn{2}{c|}{\textbf{Dimensional}} \\
\hline
& \textbf{Syndrome-specific} & \textbf{Transdiagnostic} & \textbf{Syndrome-specific} & \textbf{Transdiagnostic} \\
\hline
\textbf{Not validated} & r/SocialAnxiety membership status (member vs. not member) \newline\newline Twitter-disclosed social anxiety disorder diagnosis (present vs. absent) \newline\newline Self-reported history of social anxiety disorder diagnosis & r/Anxiety membership status (member vs. not member) \newline\newline Twitter-disclosed anxiety (present vs. absent) & Self-reported social anxiety disorder severity (e.g., ``How much have you been bothered by social anxiety disorder symptoms in the past month, on a 0-10 scale?'') & Self-reported anxiety severity (e.g., ``How much have you been bothered by anxiety symptoms in the past month, on a 0-10 scale?'') \\
\hline
\textbf{Validated} & Clinician-rated SCID-5 social anxiety disorder status (present vs. absent) \newline\newline  Self-reported SPIN score, converted to dichotomous variable reflecting the presence (score > 19) or absence (score $\leq$ 18) of social anxiety disorder & Self-reported MASQ Anxious Symptoms subscale, dichotomized (``high anxious symptoms'' vs. ``low anxious symptoms'') \newline\newline Clinician-rated IMAS fear subfactor score, converted to dichotomous variable reflecting high fear or low fear & Clinician-rated ADIS-5L social anxiety disorder severity on 0 (\textit{none}) to 8 (\textit{very severe}) scale. \newline\newline Self-reported SPIN (0-68 score) \newline\newline Self-reported IDAS-II Social Anxiety Scale & Self-reported MASQ Anxious Symptoms subscale \newline\newline Clinician-rated IMAS fear subfactor \\
\hline
\end{tabular}
}
\caption{
ADIS-5L = Anxiety and Related Disorders Interview
Schedule for \textit{DSM-5} \cite{brown_anxiety_2014}; 
IDAS-II = Inventory of Depression and Anxiety Symptoms \cite{watson2012development}; 
IMAS = Interview for Mood and Anxiety Symptoms \cite{Kotov_Perlman_Gámez_Watson_2015}; 
MASQ = Mood and Anxiety Symptom Questionnaire \cite{watson1995testing}; 
SCID-5 = Structured Clinical Interview for \textit{DSM-5} \cite{first_scid_2016}; 
SPIN = Social Phobia Inventory \cite{connor2000psychometric}.} 

\label{tab:anxiety-measures}
\end{table*}

\end{document}